\documentclass[%
 reprint,
 amsmath,
 amssymb,
 pra,
]{revtex4-2}

\usepackage{dcolumn}
\usepackage{bm}
\usepackage{mathtools}
\usepackage{amsthm} 
\usepackage{color}

\usepackage{graphicx} 
\usepackage{float} 
\usepackage{hyperref}
\usepackage{siunitx}

\usepackage[labelformat=simple]{subcaption} 

\begin{document}
\title{Grating-Induced Slow-Light Enhancement of Second Harmonic Generation in Periodically Poled Crystals}
\author{Thomas E.~Maybour}
\author{Devin H.~Smith}
\author{Peter Horak}
\affiliation{Optoelectronics Research Centre, University of Southampton, Southampton~SO17~1BJ, UK}


\begin{abstract}

\indent 
The effect of slow light on second harmonic generation in a periodically poled $\chi^{(2)}$ nonlinear medium is investigated theoretically. A linear $\pi$ phase shifted grating is used to slow the group velocity of the fundamental frequency and the resulting field enhancement greatly increases the second harmonic conversion efficiency. A second linear grating at the input end ensures that all output is in the forward direction. We show that almost 100\% conversion efficiency can be achieved for continuous wave pumping at low intensities that generate negligible conversion in the absence of the slow-light grating.
\end{abstract}

\maketitle
%

\section{\label{sec:level1}Introduction}

Second-harmonic-generation (SHG) is a second order nonlinear process induced by the $\chi^{(2)}$ susceptibility tensor of a material, typically a crystal, which converts an electromagnetic wave of frequency $\omega$ into a wave at the second harmonic frequency $2\omega$ \cite{PhysRevLett.7.118,PhysRev.128.606}. 
Because of the small value of $\chi^{(2)}$ of common nonlinear materials either long device lengths or high light intensities are required to achieve efficient SHG. 

One option to reduce the pump intensity requirements is to enhance the field intensity by enclosing a nonlinear crystal within a Fabry–Perot cavity which is resonant with either the fundamental or second harmonic frequency \cite{1074007}. An experimental demonstration showed that this method increased SHG by 13\% \cite{Kozlovsky:87}. Another approach is to enhance the field intensity by using a slow light resonance: here the chromatic dispersion of a strong, narrowband resonance of either the material itself or of an appropriate waveguide structure creates a strong reduction of group velocity. A light pulse entering such a device experiences pulse compression and correspondingly produces the field enhancement necessary for enhanced SHG. 

There are many different approaches to generating slow light, for example using electromagnetically induced transparency \cite{Hau1999} or Brillouin scattering \cite{PhysRevLett.94.153902}, but here we are principally interested in slowing light with Bragg gratings. A Bragg grating \cite{RamanKash} consisting of a periodic modulation of the refractive index reflects light within a certain frequency band. At the edges of this reflection band the grating creates strong chromatic dispersion and group velocity reduction, i.e., slow light that could be used for SHG enhancement. However, strong group velocity dispersion (GVD) also leads to significant pulse broadening in this case, thereby counteracting the field enhancement.

This pulse broadening can largely by avoided by using more complex, superstructure gratings. In particular, inserting periodic phase shifts into a standard Bragg grating \cite{pigrating}, a so-called $\pi$ phase shifted grating, opens up a narrow transmission band within the stop band which permits the generation of slow-light field enhancement with zero GVD at its center \cite{PSBG}. The same effect can also be achieved by the superposition of two Bragg gratings of similar but different resonant wavelengths, a so-called moir\'e grating \cite{MFG}.

Another important factor affecting SHG conversion efficiency is phase matching between the fundamental and second harmonic waves: chromatic dispersion of the material typically leads to dephasing and thus a periodic exchange of energy between fundamental and harmonic wave instead of a continuous increase of second harmonic energy along the propagation direction. Among the different techniques that can be used to achieve phase matching, the most popular approaches are to use either a birefringent nonlinear crystal or to employ quasi-phase-matching (QPM) \cite{PhysRev.127.1918}. This last technique works by periodically modulating the sign of the $\chi^{(2)}$ susceptibility to compensate for the phase mismatch acquired between fundamental and harmonic wave during propagation. 
Phase matching can also be achieved by tailoring the dispersion of a linear grating \cite{doi:10.1063/1.1653278,1077381,BELYAKOV198194}.

For slow-light enhancement of SHG we therefore require linear gratings in a $\chi^{(2)}$ medium. While there have been theoretical studies of linear gratings with a quadratic nonlinearity \cite{PhysRevLett.78.2341,Picciau96,MARTORELL1994319,Steel:96}, it has traditionally been challenging to write linear gratings in bulk $\chi^{(2)}$ media \cite{Hukriede_2003,Kroesen:14}. However, progress has been made in producing linear gratings with high index contrast in thin-film lithium niobate \cite{Baghban:17}.  There has also been recent demonstrations of producing $\pi$ phase shift gratings in thin-film lithium niobate \cite{8749304,9193812}.  

In this work we examine the continuous wave (CW) enhancement of second harmonic generation in a QPM device by including a $\pi$-phase shifted grating tuned to the fundamental wave frequency such that this pump field experiences slow down and thus field enhancement. However, the superstructure grating achieves the slow light effect by coupling the fundamental wave into forward and backward modes which in turn generates forward and backward second harmonic modes. A second linear Bragg grating is therefore added at the input end of the device to reflect the backward second harmonic mode and thus to ensure unidirectional forward propagating output of the second harmonic. We demonstrate that a device of this type is capable of generating considerably higher second harmonic conversion efficiency compared to a standard QPM device at lower intensities and we investigate the dependence of the conversion efficiency on the device parameters.


\section{\label{sec:level2}Theoretical Model}

\begin{figure}[tb]
	\centering
	\begin{subfigure}{0.4\textwidth} 
		\includegraphics[width=\textwidth]{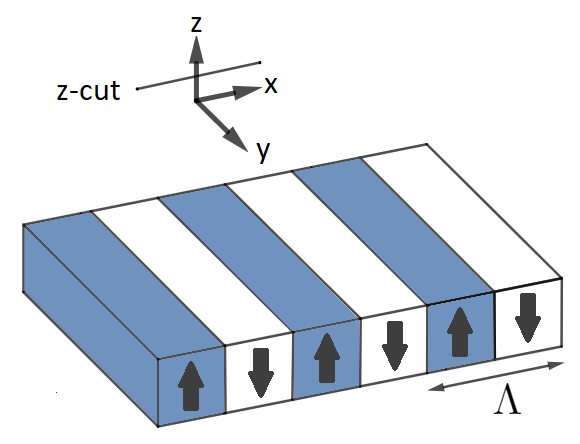}
		\caption{} 
		\label{fig:pp_medium}
	\end{subfigure}
	\vspace{1em} 
	\begin{subfigure}{0.4\textwidth} 
		\includegraphics[width=\textwidth]{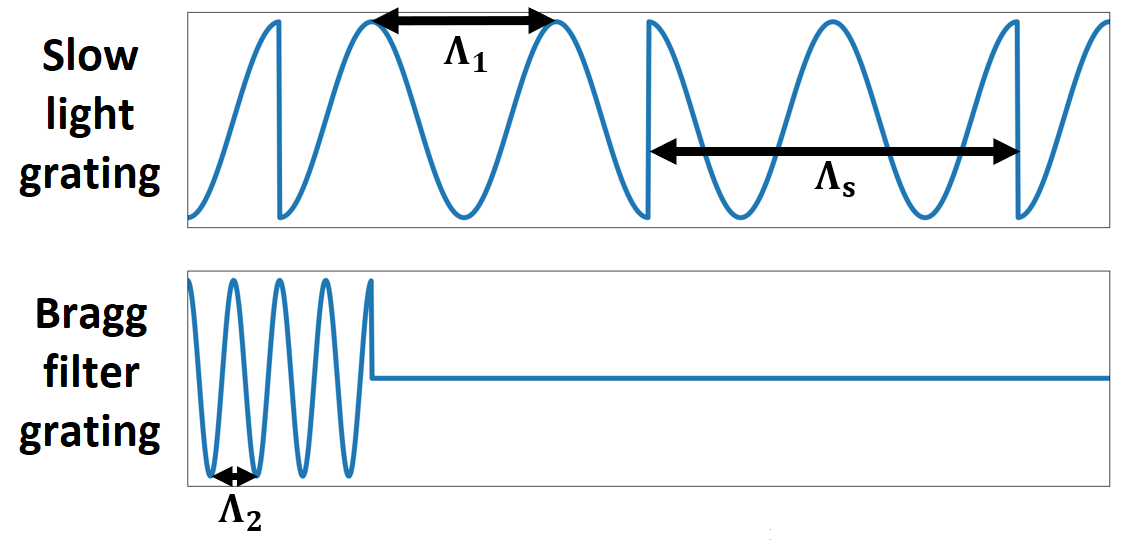}
		\caption{} 
        \label{fig:grating_schem}		
	\end{subfigure}
    \caption{(a) Schematic of a periodically poled $\chi^{(2)}$ medium showing the poling period $\Lambda$. (b) Schematic of the two linear gratings: a slow light $\pi$ phase shifted grating with Bragg period $\Lambda_{1}$ and superstructure period $\Lambda_{S}$, and a reflection Bragg grating with period $\Lambda_{2}$ at the input end of the device.}
\end{figure}

Throughout this paper we consider a device fabricated in thin film lithium niobate doped with magnesium oxide (MgO:LiNbO$_3$) to increase the optical damage threshold \cite{doi:10.1063/1.94946}. The highest nonlinear $\chi^{(2)}$ tensor component for MgO:LiNbO$_3$ is $d_{33}= \SI{25}{pm/V}$ at a wavelength of \SI{1064}{nm}, which we use in our analysis. The $d_{33}$ component is accessed by waves polarized vertically along the $z$-axis and therefore a $z$-cut thin film is required; light is propagating in the $x$ direction. The waveguide is periodically poled with period $\Lambda$ to ensure quasi-phase matching for SHG as shown in Figure \ref{fig:pp_medium}. 

In addition to the periodic variation of the $\chi^{(2)}$ nonlinearity, we assume that a spatial profile of the waveguide's linear refractive index is written into the thin film after the poling process of the form
\begin{equation}\label{refractive_index}
     n(x) = \bar{n} + \delta n \Big(f_{1}(x)a_{1}(x)a_{S}(x) + f_{2}(x)a_{2}(x)\Big),
\end{equation}
where $\bar{n}$ is the effective refractive index and $\delta n$ is the grating strength. This linear grating modulation is composed of two parts. The first part, given by the term $f_{1}(x)a_{S}(x)a_{1}(x)$, creates an apodized $\pi$-phase shifted grating with a transmission band centered at the wavelength $\lambda_{1}$ which we refer to as the slow-light grating. The second part is given by $f_{2}(x)a_{2}(x)$ and creates a Bragg reflector at the wavelength $\lambda_{2}$ of the second harmonic at the input end of the device; we refer to this term as the reflection grating. Figure \ref{fig:grating_schem} shows a schematic of these gratings.
The slow-light grating is composed of an apodization $f_{1}$ which provides an overall amplitude profile, a superstructure envelope $a_{S}$ which defines the $\pi$-phase shift positions, and a fundamental Bragg grating profile $a_{1}$. The apodization is chosen throughout the rest of this paper to have a Gaussian profile of the form 
\begin{equation}\label{apodization}
    f_{1}(x) = \exp[-\alpha_{A}(x-L/2)^{2}/L^{2}]
\end{equation}
where the center of the Gaussian is at $L/2$  and where $\alpha_{A}$ parametrizes the width of the Gaussian. In all the following analyses we set $\alpha_{A}=16$ which gives a full width at half maximum (FWHM) of $L/2\sqrt{\ln(2)}$. The superstructure envelope and the fundamental Bragg grating are given by
\begin{align}
    a_{S}(x) &= \text{sgn}\Bigg[\cos\bigg(\frac{\pi(2x-L)}{\Lambda_{s}}+\frac{\pi}{2}\bigg)\Bigg],\label{ss_env}\\
    a_{1}(x) &= \cos\bigg(\frac{2\pi x}{\Lambda_{1}} + \phi_{1}\bigg),\label{bragg1_env}
\end{align}
respectively. Here $\phi_{1}$ is a constant phase term and the fundamental Bragg period is given by $\Lambda_{1} = \lambda_{1}/(2\bar{n}_{1})$ where $\bar{n}_{1}$ is the effective refractive index at frequency $\omega$, so that the Bragg resonance is centered at the fundamental wavelength. The phase of the superstructure envelope $\pi(2z-L)/\Lambda_{s}+\pi/2$ is chosen so that there is a $\pi$ phase shift at the center of the grating for any choice of superstructure period $\Lambda_s$. 

\begin{figure}[tb]
    \includegraphics[scale=.15]{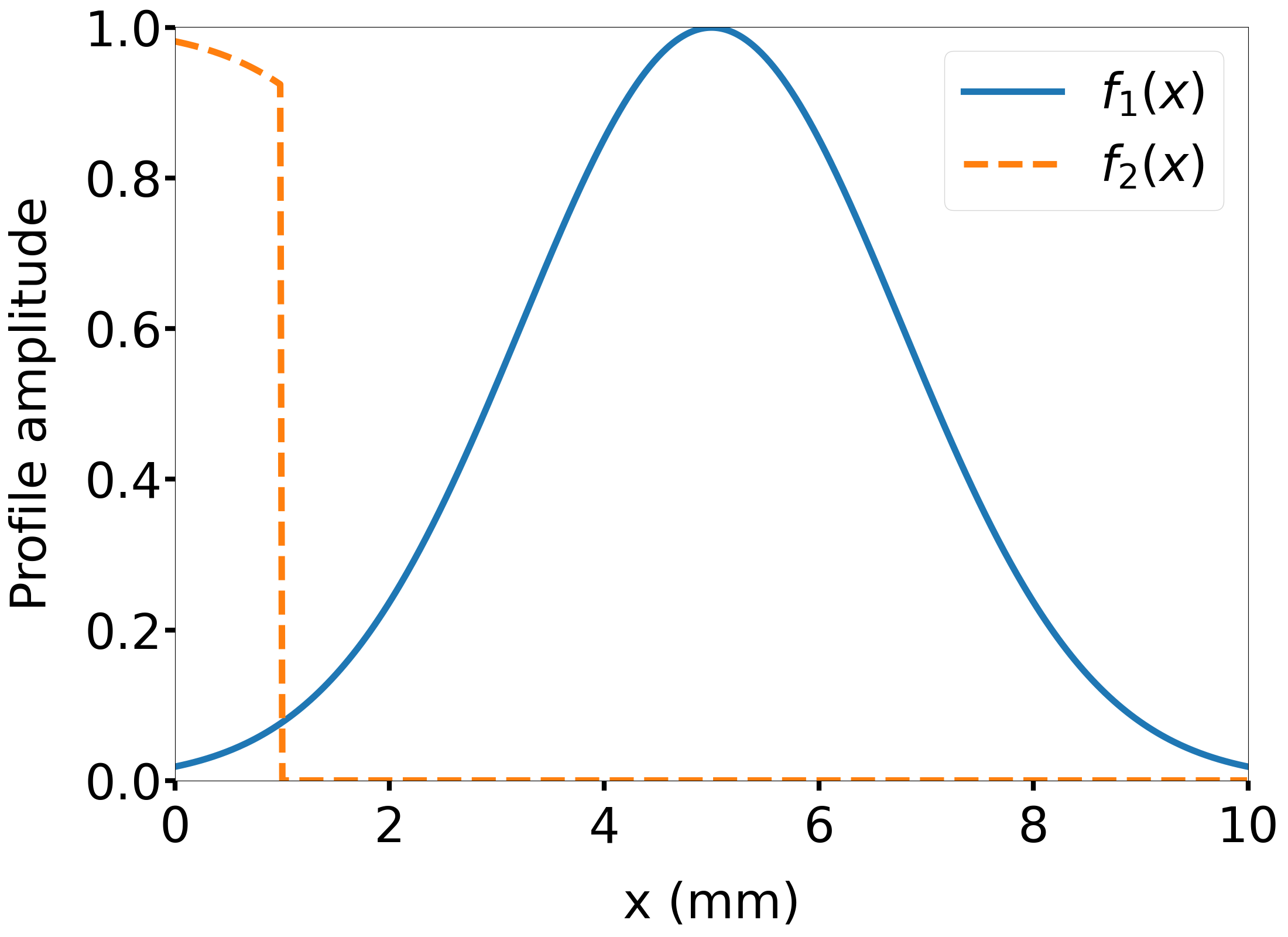}
    \centering
    \caption{Slow-light grating apodization $f_{1}$ and reflection grating profile $f_{2}$ with parameters $L=10$ mm, $\alpha_{A} = 16$ and $L_{R}=1$ mm.}
    \label{fig:apod}
\end{figure}

The reflection grating is defined by $f_{2}(x)$ which gives its overall profile and $a_{2}(x)$ which creates a Bragg grating resonant at the second harmonic. The latter is given by 
\begin{equation}
    a_{2}(x) = \cos\bigg(\frac{2\pi x}{\Lambda_{2}} + \phi_{2}\bigg),\label{bragg2_env}
\end{equation}
where $\phi_{2}$ is a constant phase term and the harmonic Bragg period $\Lambda_{2} = \lambda_{2}/(2\bar{n}_{2})$ which creates a Bragg resonance at $\lambda_{2}$ and where $\bar{n}_{2}$ is the effective refractive index at frequency $2\omega$. The apodization functions fulfil the constraint $f_1(x) + f_2(x) \leq
1$ to ensure that the overall magnitude of grating modulation does not exceed the maximum $\delta n$ that can be fabricated. Since $f_{1}(x)$ is given by Eq.\ (\ref{apodization}), we define the reflection grating profile by
\begin{equation}
    f_{2}(x) = \begin{cases} 
        1 - f_{1}(x) \, &\text{if} \quad 0 \leq x \leq L_{R},\\
        0 \, &\text{if} \quad x > L_{R},
    \end{cases}    
\end{equation}
where $ L_{R}$ is the length of the grating. Figure \ref{fig:apod} gives an example of the slow-light grating apodization and of the reflection grating profiles. 

We model light propagating through our $\chi^{(2)}$ medium with linear refractive index profile \eqref{refractive_index} by using coupled mode theory. We start with a linearly $z$-polarized electric field of the form
\begin{equation}\label{efield_ansat}
  E_{z}(x) = E_{\omega}(x) + E_{2\omega}(x)
\end{equation}
which is composed of an electric field $E_{\omega}(x)$ for the fundamental mode and $E_{2\omega}(x)$ for the second harmonic. The linear gratings will couple forward and backward propagating waves in both the fundamental and second harmonic and so we introduce the following ansatz for the fields: 
\begin{align}\label{efield_ansat_omega}
 E_{\omega}(x) &= u_{1}(x)e^{i(\beta_{1}x - \omega t)} + v_{1}(x)e^{-i(\beta_{1}x + \omega t)} + \text{c.c.},\\
\label{efield_ansat_2omega}
 E_{2\omega}(x) &= u_{2}(x)e^{i(\beta_{2}x - 2\omega t)} + v_{2}(x)e^{-i(\beta_{2}x + 2\omega t)} + \text{c.c.}
\end{align}
The forward and backward mode envelopes are given by $u_{1}$ ($u_{2}$) and $v_{1}$ ($v_{2}$), respectively, for the fundamental (harmonic) field. The propagation constants are $\beta_{1} = \bar{n}_{1}k_{1}$ $\beta_{2} = \bar{n}_{2}k_{2}$ where $k_{1}$ and $k_{1}$ are the corresponding wavenumbers for frequencies $\omega$ and $2\omega$, respectively. Coupled mode equations can be derived by substituting equations  \eqref{refractive_index} and \eqref{efield_ansat} into the nonlinear wave equation
\begin{equation}\label{wave_eq_nonlin}
    \frac{\partial^{2}E_{z}}{\partial x^{2}} - \frac{n^{2}}{c^{2}}\frac{\partial^{2}E_{z}}{\partial t^{2}} = \mu_{0}\frac{\partial^{2}P^{\text{NL} }}{\partial t^{2}}
\end{equation}
where the nonlinear polarization is given by
\begin{equation}\label{nonlin_polar}
    P^{\text{NL} } = \epsilon_{0}\chi^{(2)}(x)E_{z}^{2}. 
\end{equation}
Then by setting $\chi^{(2)}(x) = \chi^{(2)}\text{sgn}\Big[\sin(2\pi x/\Lambda)\Big]$ and making a rotating wave approximation, slowly varying envelope approximation and neglecting small terms \cite{Yariv1973}, the following set of coupled mode equations can be derived
\begin{equation}\label{cme}
\begin{aligned}
    \frac{\partial u_{1}}{\partial x} &= ie^{i\phi_{1}}\kappa_{1}(x)v_{1}+\frac{\kappa_{3}}{\bar{n}_{1}}u_{1}^{*}u_{2},\\
    \frac{\partial v_{1}}{\partial x} &= -ie^{-i\phi_{1}}\kappa_{1}(x)u_{1}+\frac{\kappa_{3}}{\bar{n}_{1}}v_{1}^{*}v_{2},\\
    \frac{\partial u_{2}}{\partial x} &= ie^{i\phi_{2}}\kappa_{2}(x)v_{2}-\frac{\kappa_{3}}{\bar{n}_{2}}u_{1}^{2},\\
    \frac{\partial v_{2}}{\partial x} &= -ie^{-i\phi_{2}}\kappa_{2}(x)u_{2}-\frac{\kappa_{3}}{\bar{n}_{2}}v_{1}^{2},\\
\end{aligned}
\end{equation}
where we introduced the coupling coefficients
\begin{align}
    \kappa_{1}(x) &= \frac{\pi\delta n}{\lambda_{1}}f_{1}(x)a_{S}(x), \\
    \kappa_{2}(x) &= \frac{2\pi\delta n}{\lambda_{1}}f_{2}(x),\\
    \kappa_{3} &= \frac{4\chi^{(2)}}{\lambda_{1}}.
\end{align}


\section{\label{sec:level3} Numerical Methods}

The coupled mode equations (\ref{cme}) form a boundary value problem with known and unknown boundary conditions on both ends of the device.
The fields have 8 complex (16 real) boundary conditions, four at the start and four at the end of the grating. The known boundary conditions at the start of the grating are $u_{1}(0) = A$ and $u_{2}(0) = 0$ where $A$ is the initial amplitude of the forward fundamental field, which is fixed by the pump intensity, and the initial forward harmonic is zero. At the end of the grating the known boundary conditions are $v_{1}(L) = 0$ and $v_{2}(L) = 0$ ensuring that no light is coupled into the gratings from the end of the structure. That leaves two boundary conditions at both the start and end of the grating that are unknown. 

Such a first order system of equations with only partially known boundary conditions can be solved numerically by the shooting method as described in detail by Ja \cite{Ja}. First we express the unknown boundary conditions by
$$
\bm{p}(x)=[v_{1}(0),v_{2}(0),u_{1}(L),u_{2}(L)]
$$ 
and express the fields by 
$$
\bm{y}(z) = [u_{1}(z),u_{2}(z),v_{1}(z),v_{2}(x)].
$$
Next an initial guess for $\bm{p}$ has to be made so that the fields  $\bm{y}(z,\bm{p})$ are now also a function of the unknown boundary conditions. We then denote integrating the fields forward from the start of the grating by $\bm{y}_{f}(x,\bm{p})$ and integrating backwards from the end of the grating by $\bm{y}_{b}(z,\bm{p})$. Then for some point $x=m$, where $m$ can be arbitrarily chosen, finding the solution to the coupled mode equations equates to solving the equation
\begin{equation}\label{shooting_eq}
    \bm{g}(\bm{p}) = \bm{y}_{f}(m,\bm{p}) - \bm{y}_{b}(m,\bm{p}) = \bm{0}.
\end{equation}

There exist numerous methods to solving Eq.~(\ref{shooting_eq}). Our approach here is as follows. As $\bm{g}(\bm{p})$ is in general a complex function we can define the quantity
\begin{equation}\label{shooting_min}
    \mathcal{L} = \sum_{i} \; \lvert  g_{i}(p_{i})\rvert^{2}
\end{equation}
and minimizing $\mathcal{L}$ to zero is equivalent to solving Eq.~(\ref{shooting_eq}). For this minimization we use the Nelder–Mead method \cite{NeldMead}. Compared to many approaches of directly solving Eq.~(\ref{shooting_eq}), this has the advantage that it does not require calculating the Jacobian and therefore the partial derivatives of $\bm{g}(\bm{p})$. In practice we have found that the convergence of the Nelder–Mead method fails for high intensities if a poor initial choice of $\bm{p}$ is made. Therefore to find solutions for higher intensities it is necessary to first find a solution that converges at a lower intensity and then incrementally increase the intensity up to the desired value. At each increment the initial choice for the unknown boundary conditions is then the $\bm{p}$ found at the previous increment.  

Our system has a number of free parameters: the superstructure period $\Lambda_{s}$, the two Bragg phases $\phi_{1}$ and $\phi_{2}$, the length of the slow-light grating $L$, the length of the reflection grating $L_{R}$, the input intensity of the forward fundamental mode $I$, and the grating strength $\delta n$.  

The aim of our study is to maximize SHG, i.e., to maximize the forward propagating second harmonic field $u_2(x=L)$ at the device output. We are thus seeking to find the parameters which achieve this aim by studying numerically the parameter dependence of the solutions of Eq.~(\ref{shooting_eq}) or, equivalently, Eq.~(\ref{shooting_min}).


\section{\label{sec:level4}Group velocity and intensity enhancement}

\begin{figure}[tb]
    \includegraphics[scale=.12]{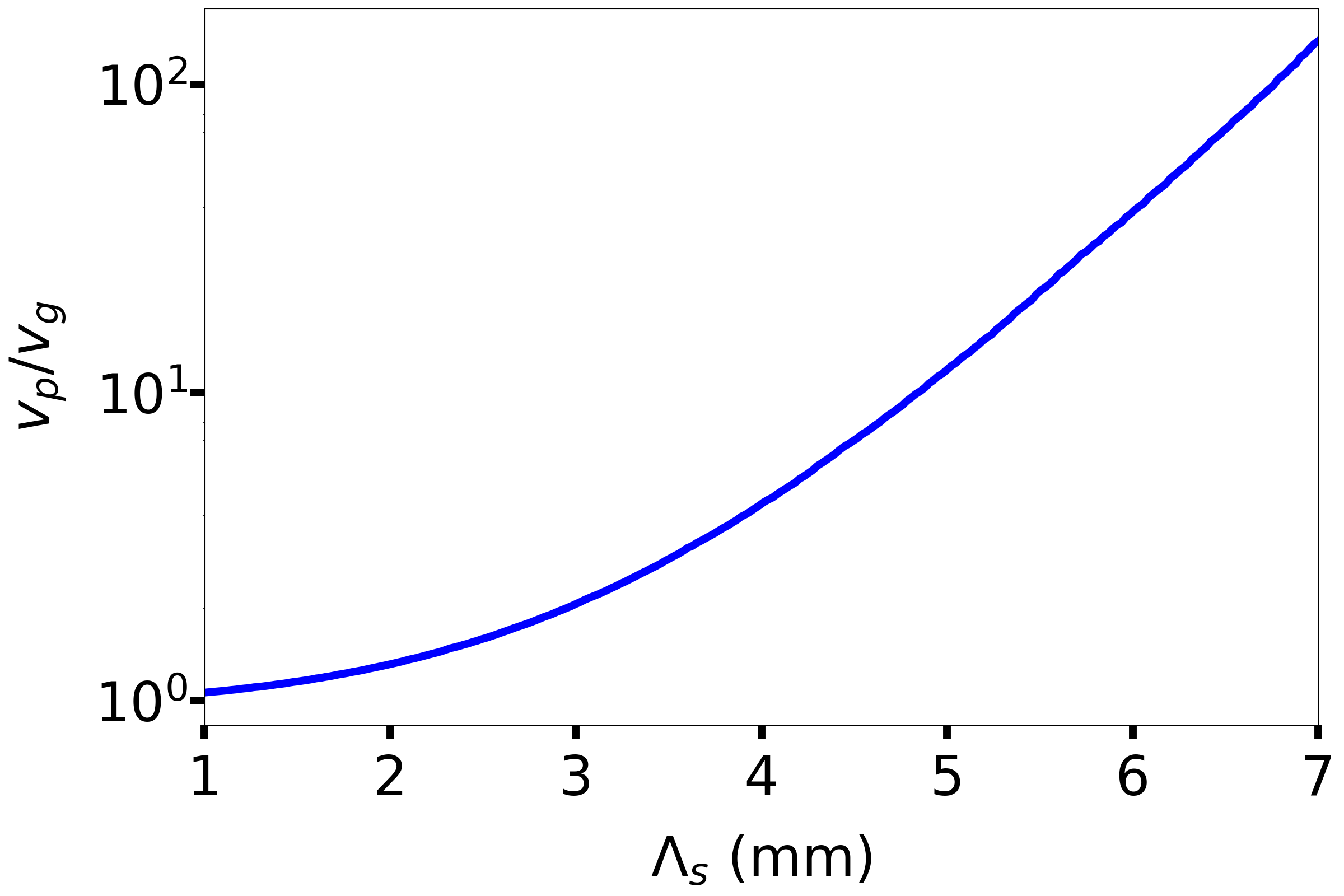}
    \centering
    \caption{Group velocity of a Gaussian apodized $\pi$-phase shifted grating vs superstructure period $\Lambda_s$. With parameters  $\alpha_{A} = 16$, $\lambda_{1} = 1064$\,nm, $\bar{n} = 2.147$, $\delta n = 10^{-3}$ and $L = 4$\,cm} 
    \label{fig:gv_log}
\end{figure}

The superstructure period $\Lambda_{s}$ is a key parameter as it controls the bandwidth of the transmission band. A longer $\Lambda_{s}$ leads to a narrower transmission band with a reduced group velocity. Typically the group velocity is defined by $v_{g}=\partial{\omega}/\partial{\beta}$, however, the propagation constant $\beta$ is modified due to the presence of the $\pi$-phase shifted grating and cannot be directly calculated. It can be shown that the group velocity can instead be defined by
\begin{equation}\label{gv}
    v_{g} = v_{p}\frac{\int_0^{L} \mathrm{d}x \; \lvert u_{1}\rvert^{2} - \lvert v_{1}\rvert^{2}}{\int_0^{L} \mathrm{d}x \; \lvert u_{1}\rvert^{2} + \lvert v_{1}\rvert^{2}},
\end{equation}
which is expressed directly in terms of the forward and backward propagating fundamental fields and where $v_{p}$ is the phase velocity \cite{PSBG,PhysRevA.104.013503}. Figure \ref{fig:gv_log} shows how the group velocity slow down factor defined by $v_{p}/v_{g}$ varies with $\Lambda_{s}$ in a Gaussian apodized $\pi$-phase shifted grating. Lower group velocities lead to greater slow down factors and field enhancement which in turn allows for more efficient SHG as discussed below.

\section{\label{sec:level5} Results}

\begin{figure}[tb]
	\centering
	\begin{subfigure}{0.4\textwidth} 
		\includegraphics[width=\textwidth]{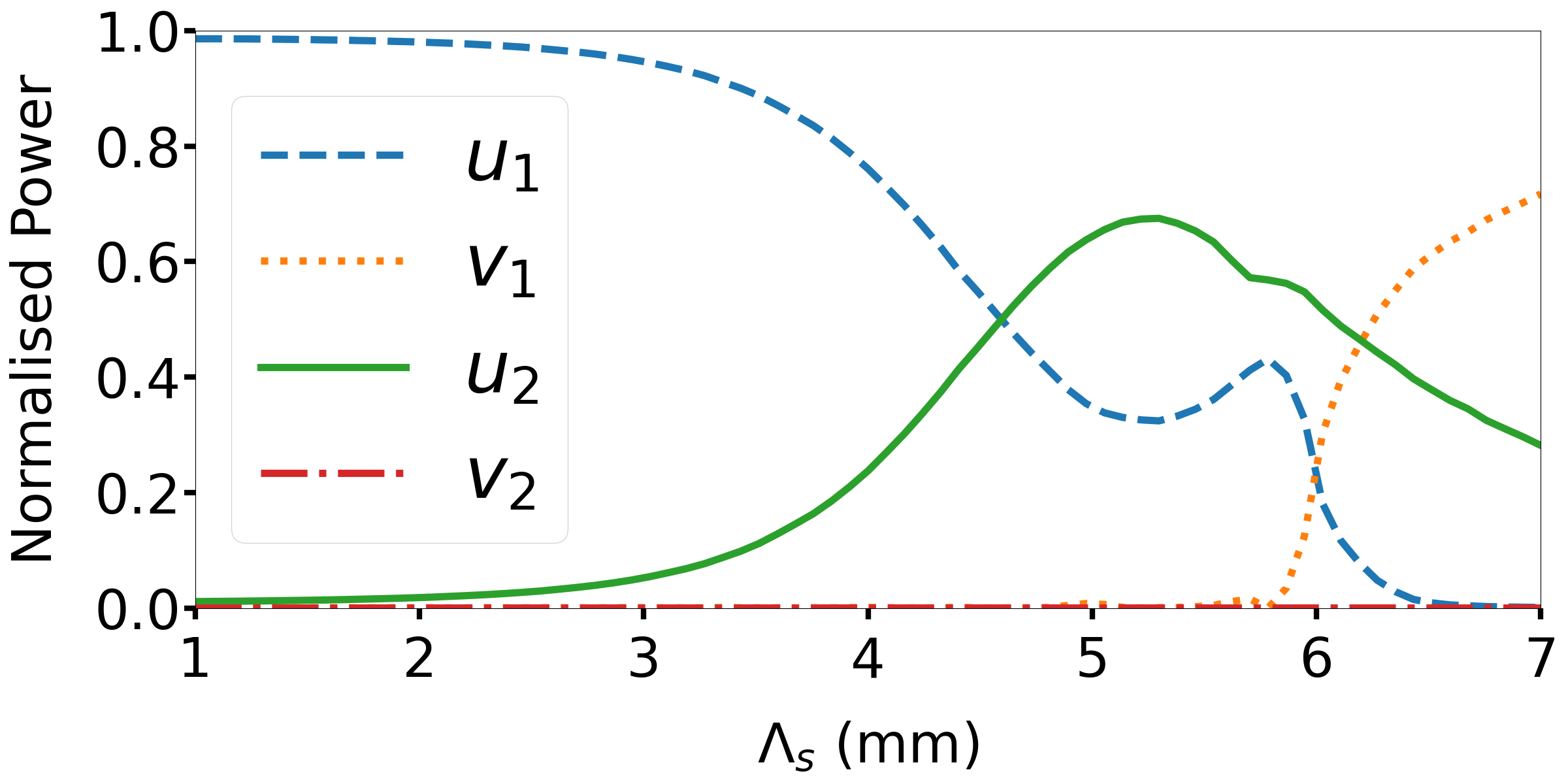}
		\caption{} 
		\label{fig:dsv_ratios}
	\end{subfigure}
	\vspace{1em} 
	\begin{subfigure}{0.4\textwidth} 
		\includegraphics[width=\textwidth]{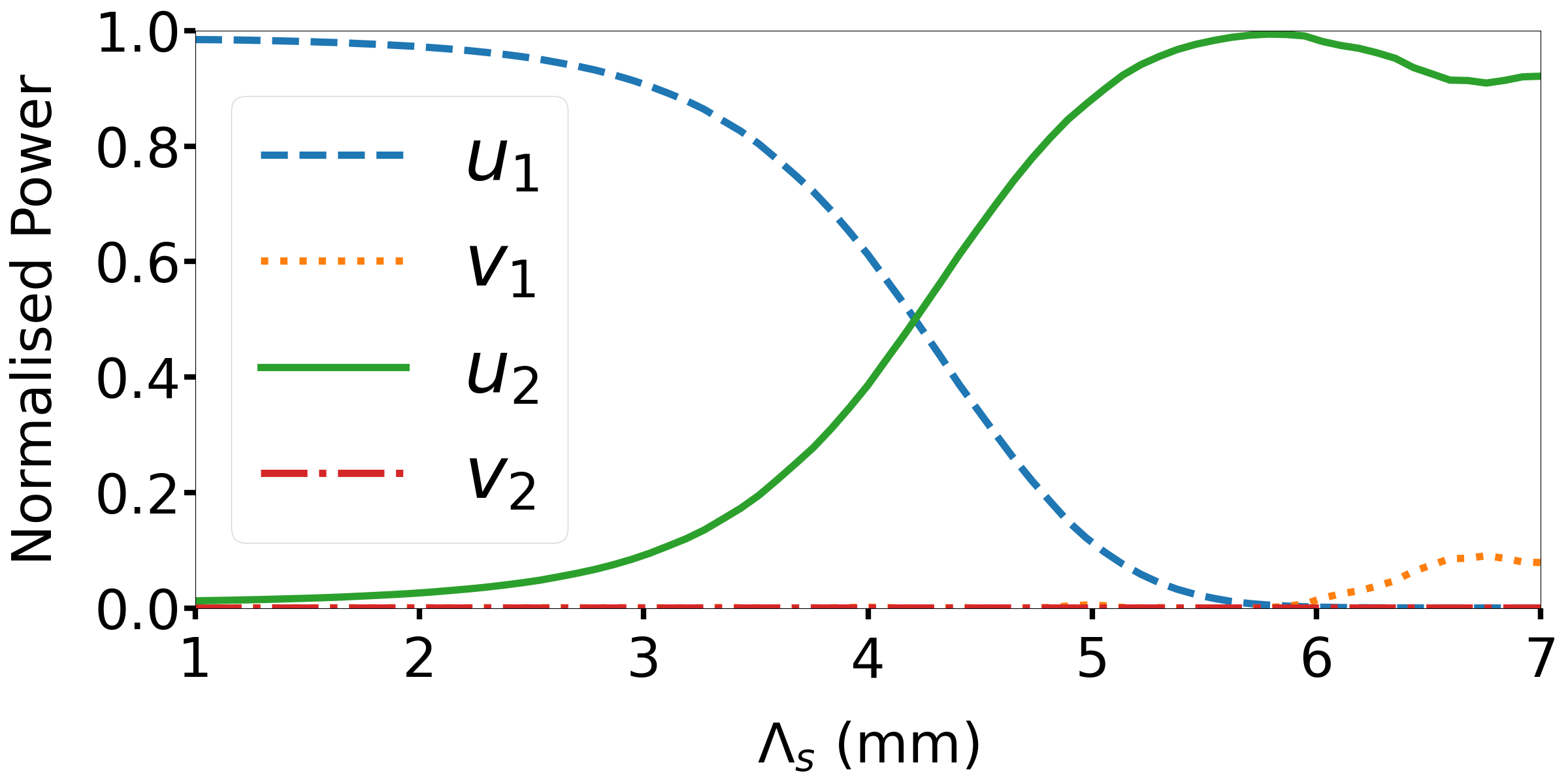}
		\caption{} 
        \label{fig:dsv_ratios_opto}		
	\end{subfigure}
	\vspace{1em} 
	\begin{subfigure}{0.4\textwidth} 
		\includegraphics[width=\textwidth]{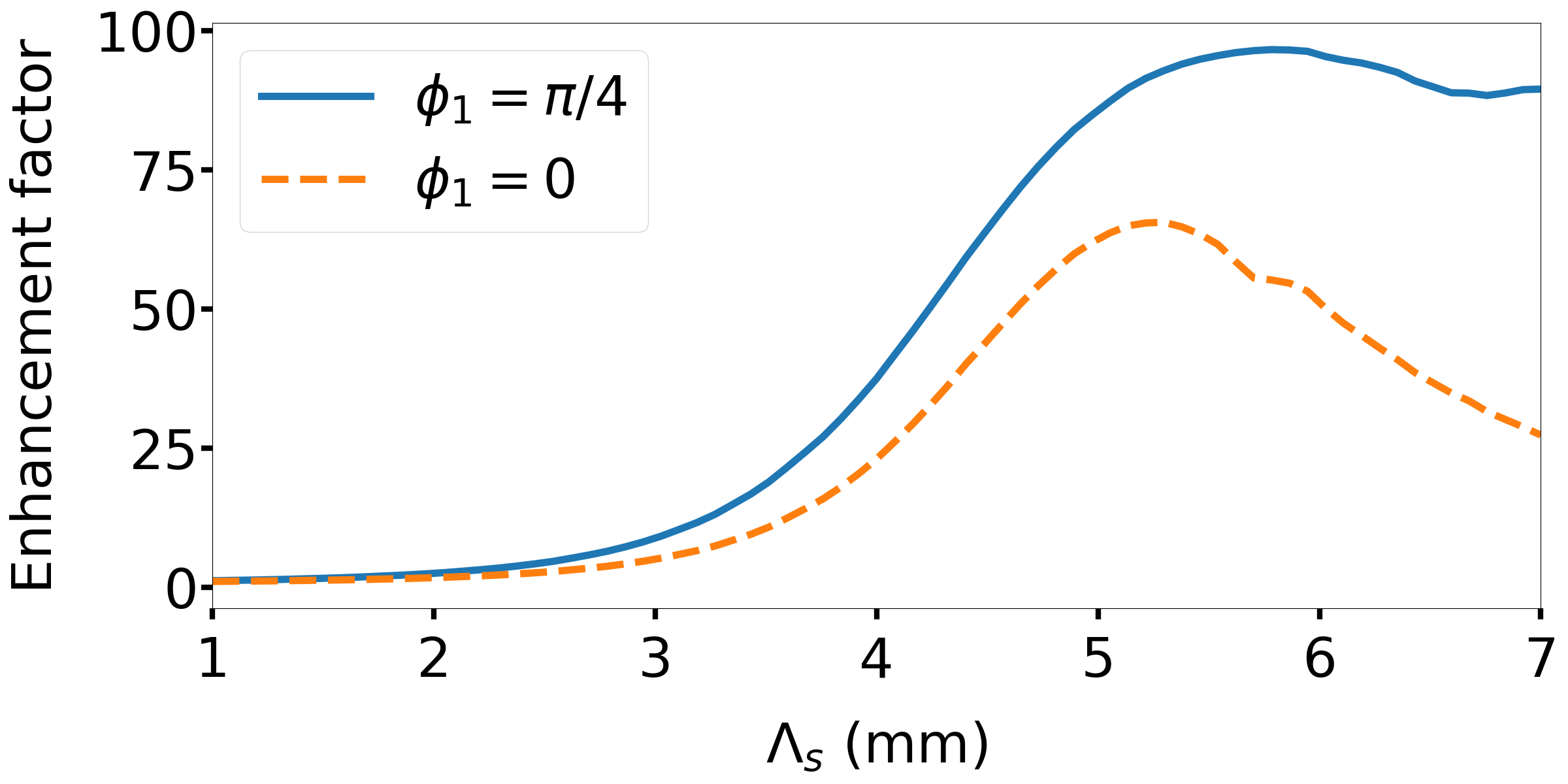}
		\caption{} 
        \label{fig:dsv_enhancement}
    \end{subfigure}	
    \caption{Output powers for fundamental and second harmonic fields versus superstructure period $\Lambda_{s}$, where powers are normalized to the input fundamental power, for (a) $\phi_{1} = 0$, and (b) $\phi_{1} = \pi/4$. (c) Enhancement of SHG conversion efficiency compared to a periodically poled crystal without linear gratings. 
    Other parameters are $\lambda_{1} = 1064$\,nm, $L = 4$\,cm, $L_{R} = 1$\,mm, $\alpha_{A} = 16$, $\delta n = 10^{-3}$, $\bar{n}_{1} = 2.147$, $\bar{n}_{2} = 2.223$, $\chi^{(2)} = 25$\,pm/V and $I = 10^{3}$\,W/c$\text{m}^{2}$. \label{fig:dsv}}	
\end{figure}

To first demonstrate the effectiveness of the slow-light and reflection gratings in enhancing second harmonic generation, figure \ref{fig:dsv_ratios} shows a simulation of the output powers for the four propagating fields, i.e., at $x=L$ for $u_1$ and $u_2$ and at $x=0$ for $v_1$ and $v_2$, when the superstructure period $\Lambda_{s}$ is varied. The other parameters were fixed to $\phi_{1}=\phi_{2}=0$, $L = 4$\,cm, $L_{R} = 1$\,mm, $I = 10^{3}$\,W/c$\text{m}^{2}$ and $\delta n = 10^{-3}$. As $\Lambda_{s}$ is increased we find a corresponding increase of the second harmonic generation. The maximum second harmonic generation occurs at $\Lambda_{s}= $ \SI{5.3}{mm} which corresponds to a slow down factor of 16.6 and a conversion efficiency of 67\% which is an enhancement by a factor of 65.5 compared to the periodically poled crystal without the linear gratings, as shown in figure \ref{fig:dsv_enhancement}.

Once $\Lambda_{s}$ is past its optimum value we find that power begins transferring to the backwards fundamental mode as can be seen in figure \ref{fig:dsv_ratios}. To understand this behavior, we note that there are two channels by which the input power of $u_{1}$ can be transferred to $u_{2}$. It can be transferred directly via the nonlinearity $\kappa_{3}$ between the forward propagating modes, or by first coupling the forward into the backward fundamental mode $v_{1}$ by the slow-light grating via $\kappa_{1}$, then into the backward harmonic mode $v_{2}$ by $\kappa_{3}$ and then finally to the forward harmonic mode $u_{2}$ via the reflection grating $\kappa_{2}$. However, depending on the relative phases of fundamental and harmonic fields, the same processes can also convert power back from the harmonic to the fundamental mode.

\begin{figure}[tb]
	\centering
	\begin{subfigure}{0.4\textwidth} 
		\includegraphics[width=\textwidth]{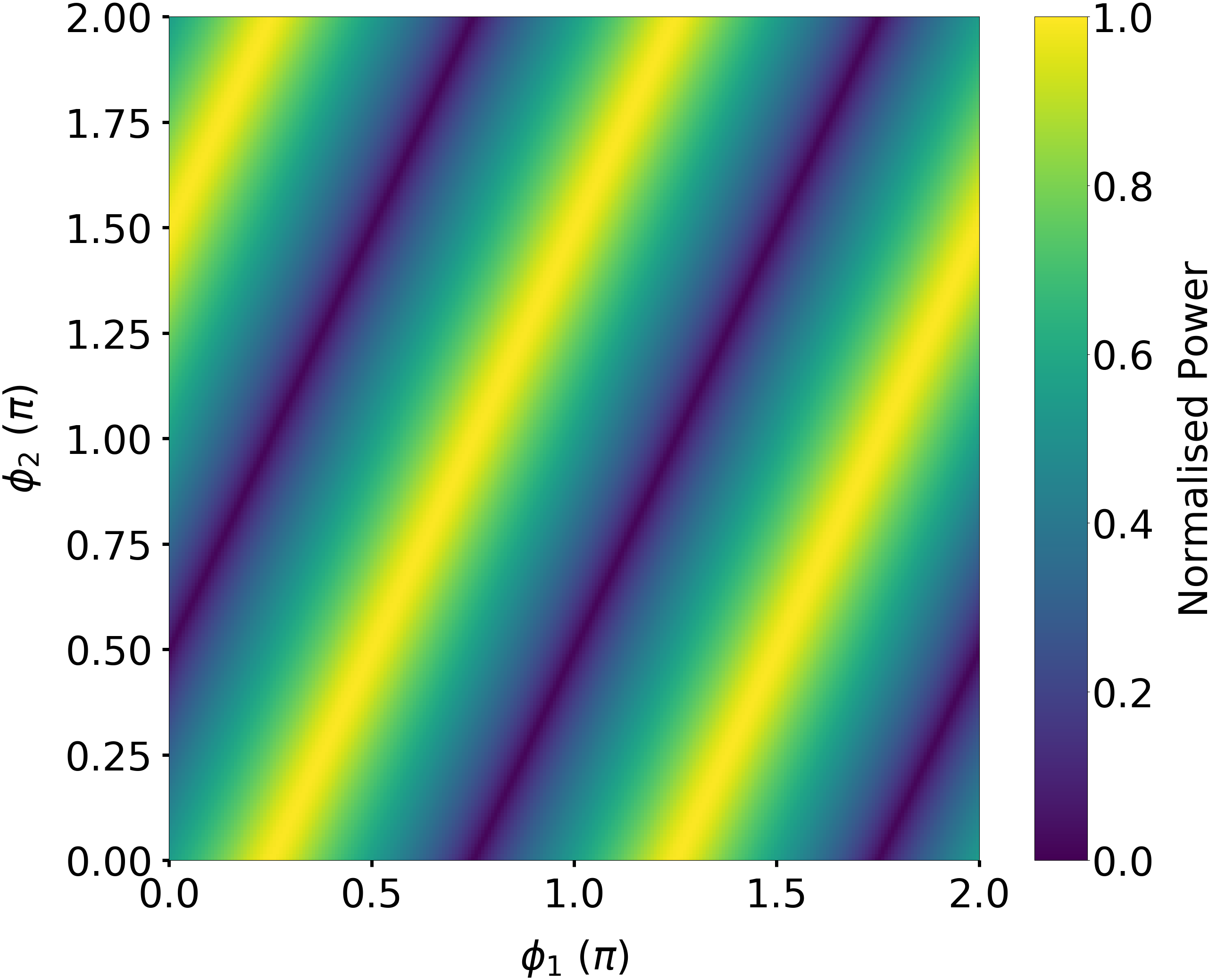}
		\caption{} 
	\end{subfigure}
	\vspace{1em} 
	\begin{subfigure}{0.4\textwidth} 
		\includegraphics[width=\textwidth]{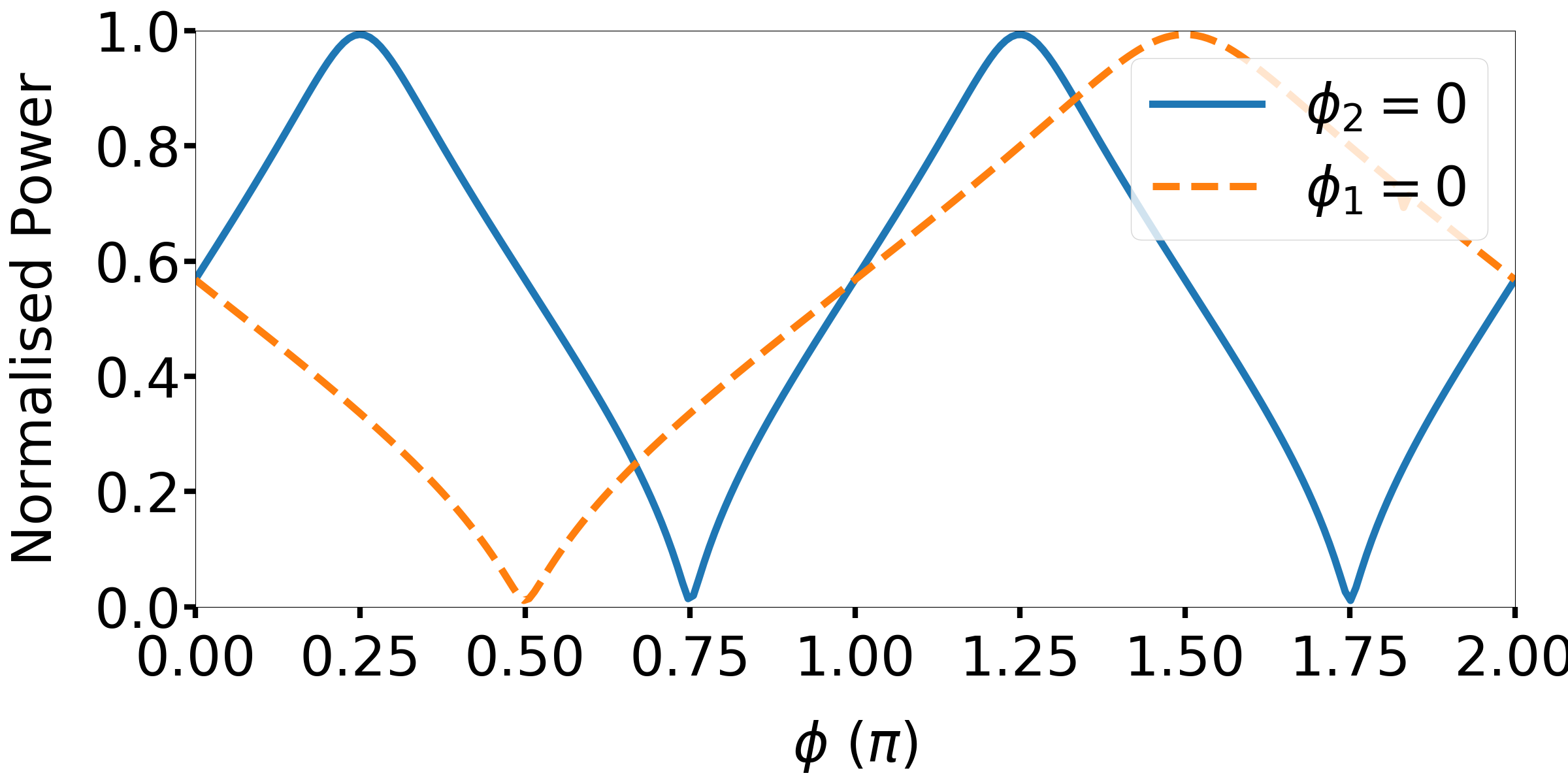}
		\caption{} 
	\end{subfigure}
    \caption{(a) Normalized output power of the forward second harmonic field versus Bragg phases $\phi_{1}$ and $\phi_{2}$. (b) 1-dimensional cuts through (a) at $\phi_1=0$ and $\phi_2=0$, respectively. $\Lambda_{s} = 5.3$\,mm, other parameters as in Fig.~\ref{fig:dsv}. \label{fig:phi1_phi2}}
\end{figure}

Therefore, we next look at the affect of varying the Bragg phases $\phi_{1}$ and $\phi_{2}$. Figure \ref{fig:phi1_phi2} shows a simulation of the normalized output power of the forward second harmonic when $\phi_{1}$ and $\phi_{2}$ are varied through $2\pi$. The figure shows that the interaction of the two linear gratings can create a resonance or antiresonance depending of the values of $\phi_{1}$ and $\phi_{2}$. This behaviour is not seen if the reflection grating is removed. We note that the condition
\begin{equation}\label{phase_constraint_max}
     \phi_{1} - \frac{\phi_{2}}{2} = \frac{\pi}{4}
\end{equation}
gives the optimum second harmonic generation. In all the simulations we have conducted Eq.~\eqref{phase_constraint_max} holds for any choice of parameters. Similarly the condition 
\begin{equation}\label{phase_constraint_min}
     \phi_{1} - \frac{\phi_{2}}{2} = \frac{3\pi}{4}
\end{equation}
gives a minimum for SHG exiting the device, with all of the power staying in the forward fundamental mode.

For such an optimized situation, $\phi_{1} = \pi/4$ and $\phi_{2} = 0$, Figure \ref{fig:dsv_ratios_opto} shows the various field output powers versus $\Lambda_s$. In this case, the optimum forward second harmonic conversion efficiency is increased to 99\% at $\Lambda_{s}=5.8$\;mm corresponding to a slow down factor of 26.2.

\begin{figure}[tb]
    \includegraphics[scale=.12]{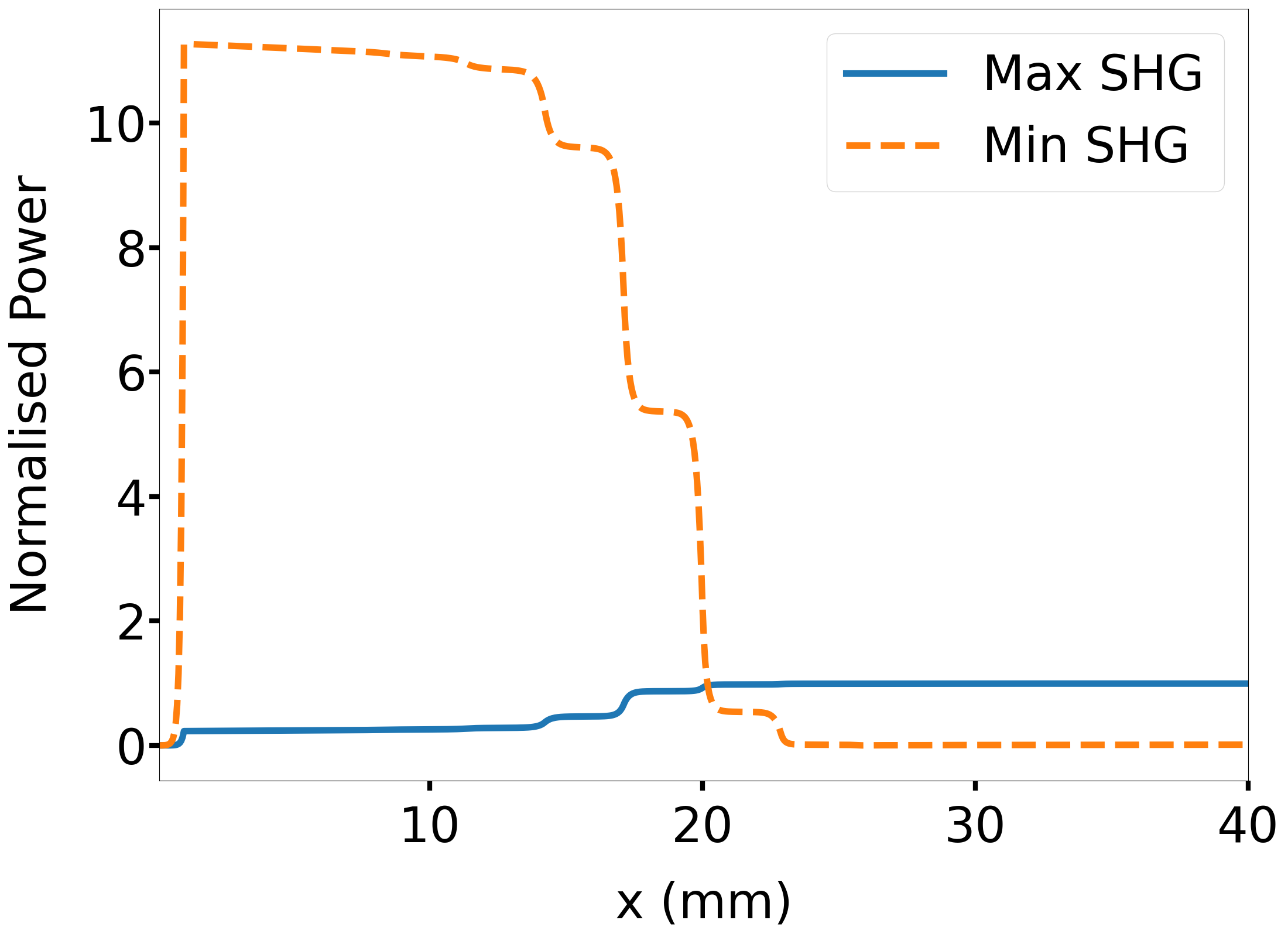}
    \centering
    \caption{Comparison of power flow normalized to input power across the device length for the forward second harmonic mode, with $\phi_{1}$ set to produce maximum and minimum conversion to the second harmonic, respectively. Here $\Lambda_{s} = 5.78$\,mm, other parameters as in Fig.~\ref{fig:dsv}(b)}
    \label{fig:max_min_shg}
\end{figure}

Finally, Figure \ref{fig:max_min_shg} shows the forward second harmonic field intensities along the length of the device with the Bragg phases set to \eqref{phase_constraint_max} and \eqref{phase_constraint_min}, respectively. We can see that in the case of Eq.~\eqref{phase_constraint_min} a strong resonator is formed between the reflection grating and the slow-light grating, where high second harmonic intensities are generated close to the reflection grating. However, as the fields propagate along $x$, this power is converted back into the fundamental wave, with notable ``steps'' at the positions of the $\pi$ phase shifts of the slow-light grating, and therefore little second harmonic output is observed at the far end of the device. For phases fulfilling Eq.~\eqref{phase_constraint_max}, on the other hand, the second harmonic intensity builds up continuously along $x$ and reaches its maximum at the device end. Therefore, the relative phase difference between the gratings has a strong affect on the overall efficiency of the second harmonic generation. 

\begin{figure}[tb]
	\centering
	\begin{subfigure}{0.4\textwidth} 
		\includegraphics[width=\textwidth]{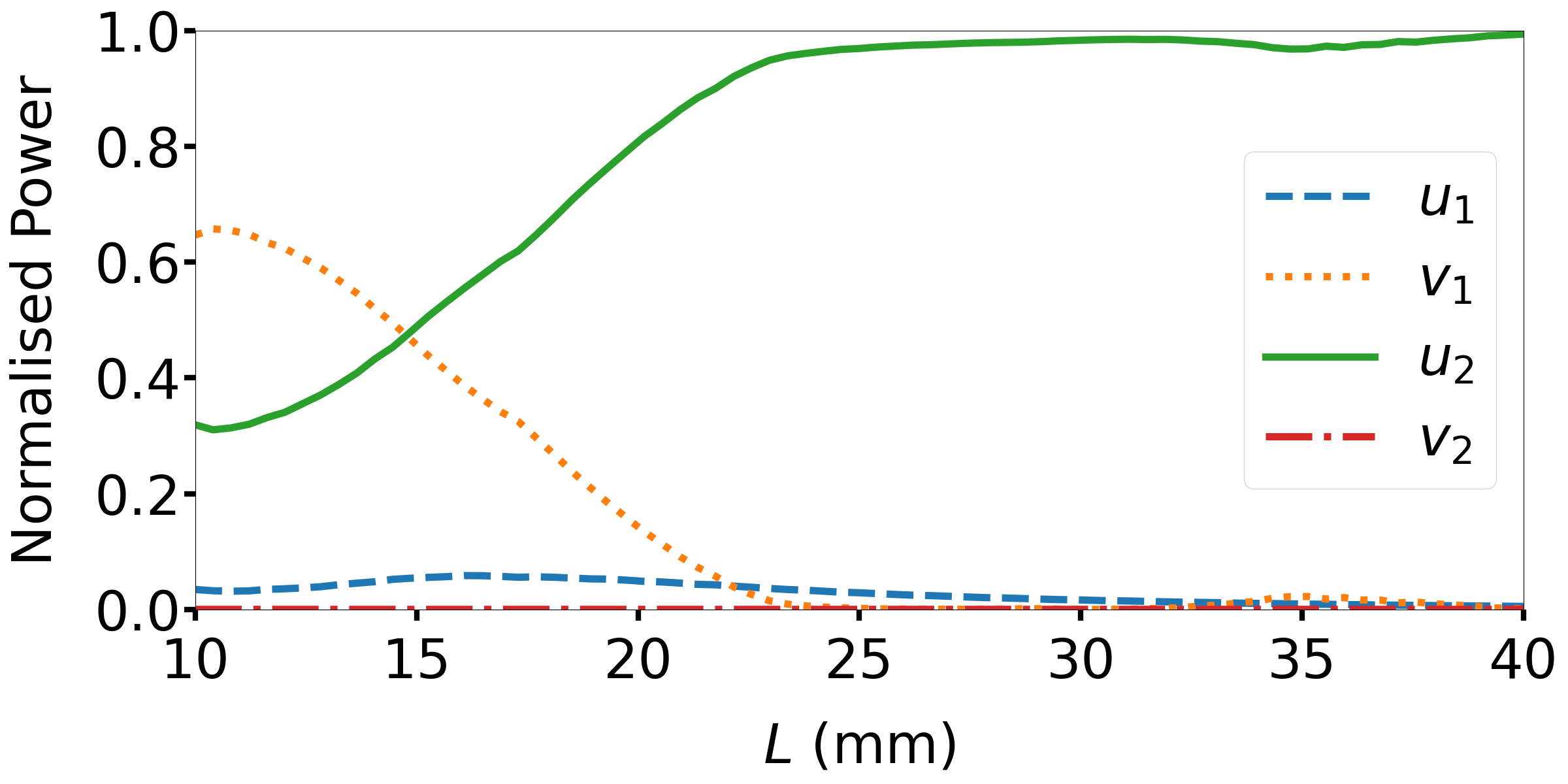}
		\caption{} 
		\label{fig:xlenv_ratios}
	\end{subfigure}
	\vspace{1em} 
	\begin{subfigure}{0.4\textwidth} 
		\includegraphics[width=\textwidth]{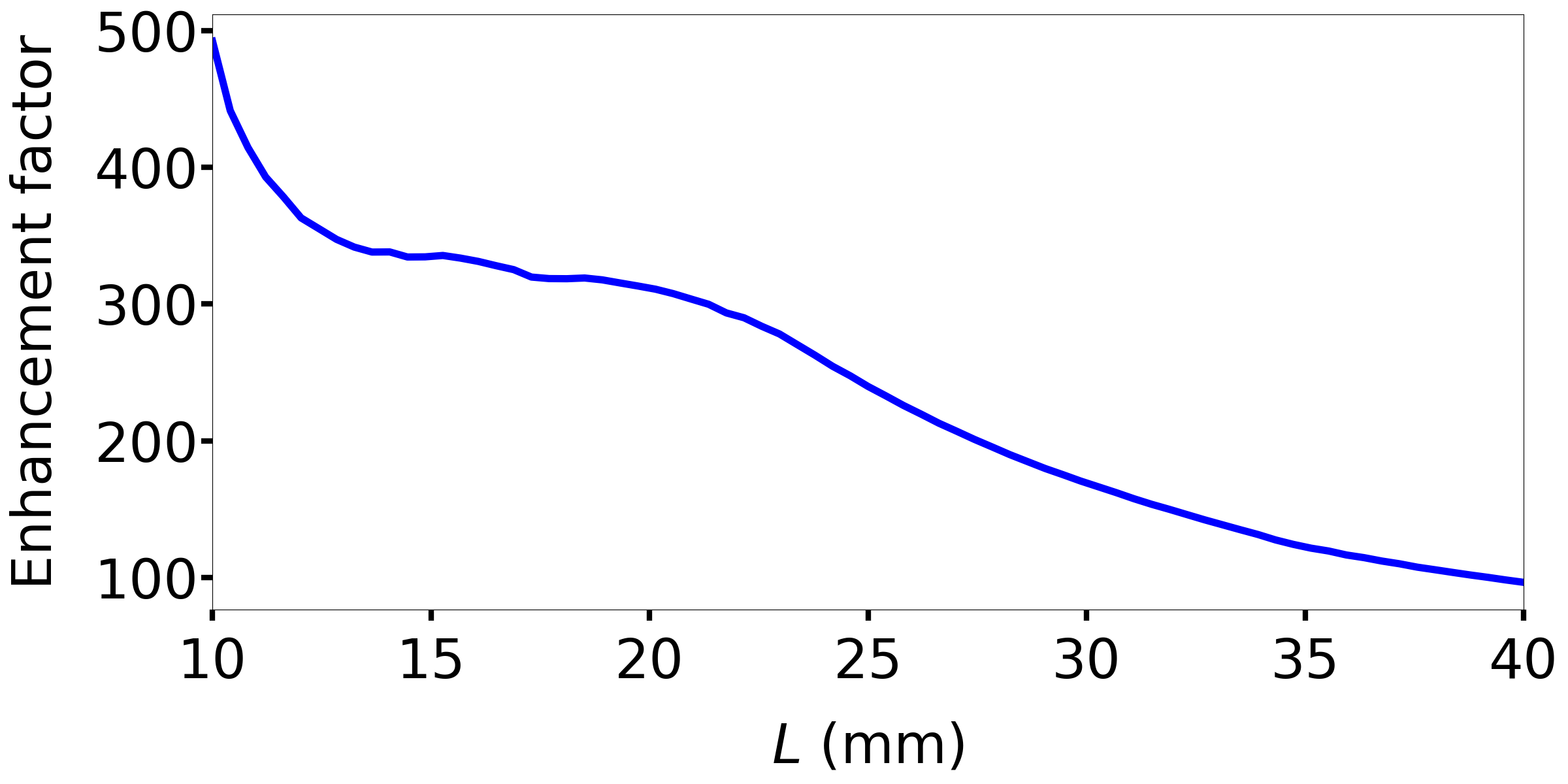}
		\caption{} 
        \label{fig:len_enhancement}		
	\end{subfigure}
	\vspace{1em} 
	\begin{subfigure}{0.4\textwidth} 
		\includegraphics[width=\textwidth]{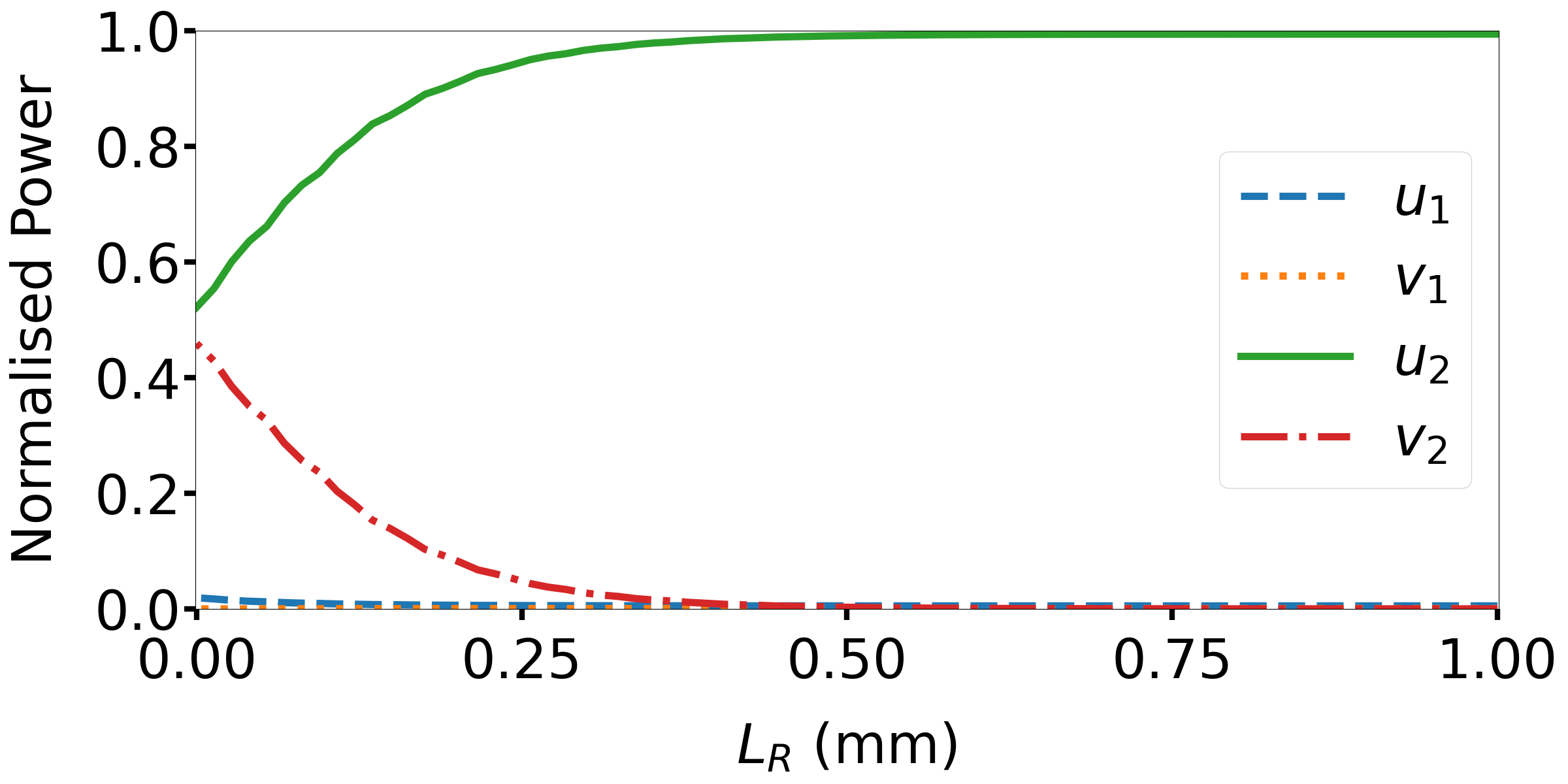}
		\caption{} 
        \label{fig:flv_ratios}		
	\end{subfigure}
    \caption{(a) Normalized output powers for fundamental and second harmonic modes versus device length $L$.\\ (b) Corresponding enhancement of SHG efficiency compared to a QPM device without linear gratings. \\
    (c) Output powers versus length of the reflection grating $L_{R}$. Here $\Lambda_{s} = 5.78$\,mm, other parameters as in Fig.~\ref{fig:dsv}(b). \label{fig:lengths}}
\end{figure}

Another factor that affects the conversion efficiency is the length of the waveguide. In a standard QPM device, the longer the interaction length the higher the conversion efficiency. The same is true for our device, as can be seen from Figure \ref{fig:xlenv_ratios} which shows the output fields as a function of device length. The parameters used here are those which we found previously to optimize the conversion efficiency for a 4\,cm device with an input intensity of $10^{3}$\,W/c$\text{m}^{2}$. The figure shows that the conversion efficiency remains close to 100\% down to a device length of around 2.5\,cm after which the efficiency starts to decline; at 10\,mm the efficiency is at 32\%. Figure \ref{fig:len_enhancement} shows the enhancement of SHG efficiency compared to a QPM device without the linear gratings. For a short, 10 mm length device, the enhancement factor is $492$ which is a significant increase in enhancement over the 4\,cm device. In a simplified picture we can argue that for the chosen value of $\Lambda_s$ the slow down factor is 26.2, c.f. Fig.~\ref{fig:gv_log}, and thus we may expect an enhancement of the fundamental wave intensity by the same factor. Since SHG scales with the square of the pump field, the SHG enhancement by the slow-light effect should be of the order of 600 which is comparable with the numerically found value. Note, however, that this simplified argument neglects depletion of the pump field and the additional field enhancements due to the resonator effect between the reflection and slow-light grating as discussed above. Thus, while the slow-light enhanced conversion efficiency converges to close to 100\% already at short device lengths, the conversion efficiency of a simple QPM device still increases quadratically with length which explains the decay of the curve in Fig.~\ref{fig:len_enhancement} for longer lengths $L$.

The length of the reflection grating is also important for increasing SHG conversion efficiency. Figure \ref{fig:flv_ratios} shows how varying the reflection grating length from 0 to 1 mm affects the SHG. When the grating is removed, $L_R=0$, the second harmonic is split almost evenly between the forward and backward outputs. As the grating length is increased we see the backward mode being converted into the forward mode. Once the grating is sufficiently long to become a near-perfect reflector, i.e., for $L_R \gg 1/\kappa_2 = 0.17$ mm, no backward harmonic light is transmitted and increasing $L_R$ further does not contribute any further to increasing the conversion efficiency.

\begin{figure}[tb]
	\centering
	\begin{subfigure}{0.4\textwidth} 
		\includegraphics[width=\textwidth]{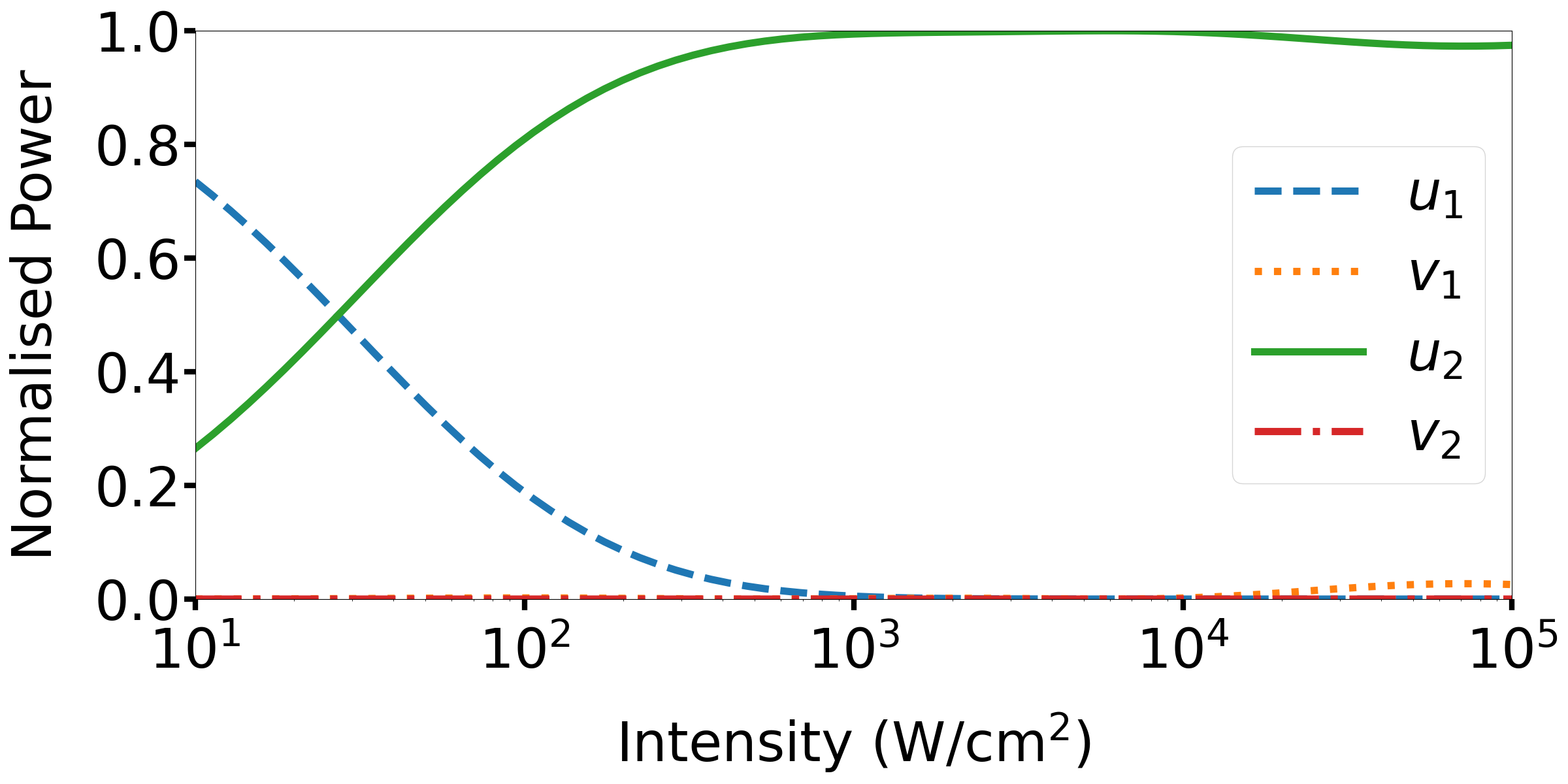}
		\caption{} 
		\label{fig:int_ratios}
	\end{subfigure}
	\vspace{1em} 
	\begin{subfigure}{0.4\textwidth} 
		\includegraphics[width=\textwidth]{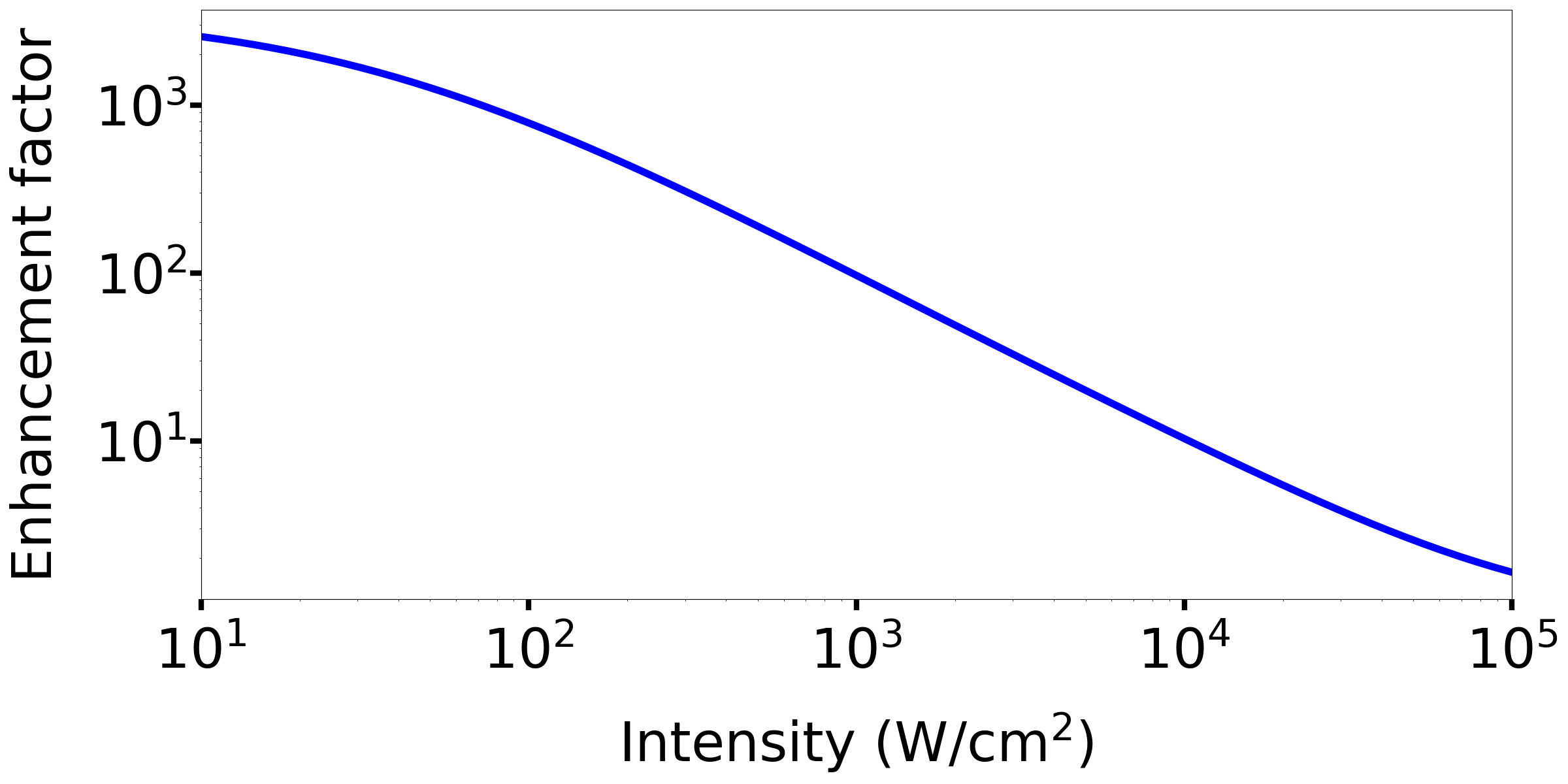}
		\caption{} 
        \label{fig:int_enhancement}		
	\end{subfigure}
    \caption{(a) Output powers for fundamental and second harmonic modes versus input intensity. \\ (b) Corresponding SHG enhancement over a standard QPM device. Here $\Lambda_{s} = 5.78$\,mm, other parameters as in Fig.~\ref{fig:dsv}(b)}
\end{figure}

So far we assumed an input intensity of $10^{3}$\,W/c$\text{m}^{2}$ which in a 4 cm standard QPM device has a low conversion efficiency and therefore is a good intensity to demonstrate the performance of our device. However, the conversion efficiency is intensity dependent and we next investigate over what range of input intensities our device remains effective. Figure \ref{fig:int_ratios} shows how the conversion efficiency varies from an intensity of $10$\,W/c$\text{m}^{2}$ to $10^{5}$\,W/c$\text{m}^{2}$ for device parameters which were optimized for an input intensity of $10^{3}$\,W/c$\text{m}^{2}$. The figure shows that for the given parameters the conversion efficiency is close to 100\% for three orders of magnitude from $10$\,W/c$\text{m}^{3}$ to $10^{5}$\,W/c$\text{m}^{5}$. Below $10^{3}$\,W/c$\text{m}^{2}$ the conversion efficiency begins to decrease, reaching 25\% efficiency at $10^{1}$\,W/c$\text{m}^{2}$ which is an enhancement factor of 2555 compared to a standard QPM device, see Fig.~\ref{fig:int_enhancement}. This is analogous to what we saw when varying the device length, lower intensities have greater enhancement factors whereas higher intensities have greater conversion efficiency. 

\begin{figure}[tb]
	\centering
	\begin{subfigure}{0.4\textwidth} 
		\includegraphics[width=\textwidth]{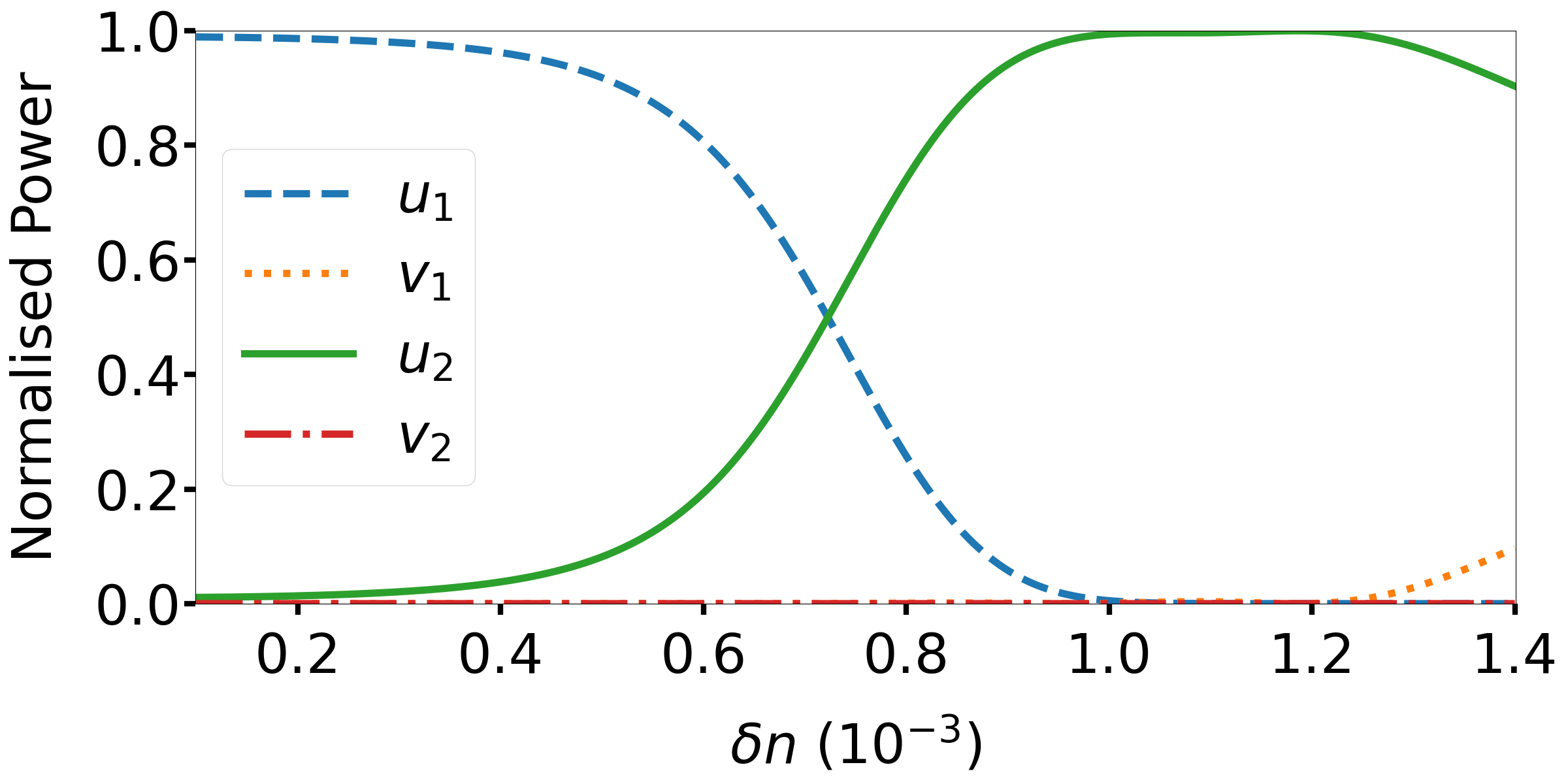}
		\caption{} 
        \label{fig:dnv_ratios}		
	\end{subfigure}
	\vspace{1em} 
	\begin{subfigure}{0.4\textwidth} 
		\includegraphics[width=\textwidth]{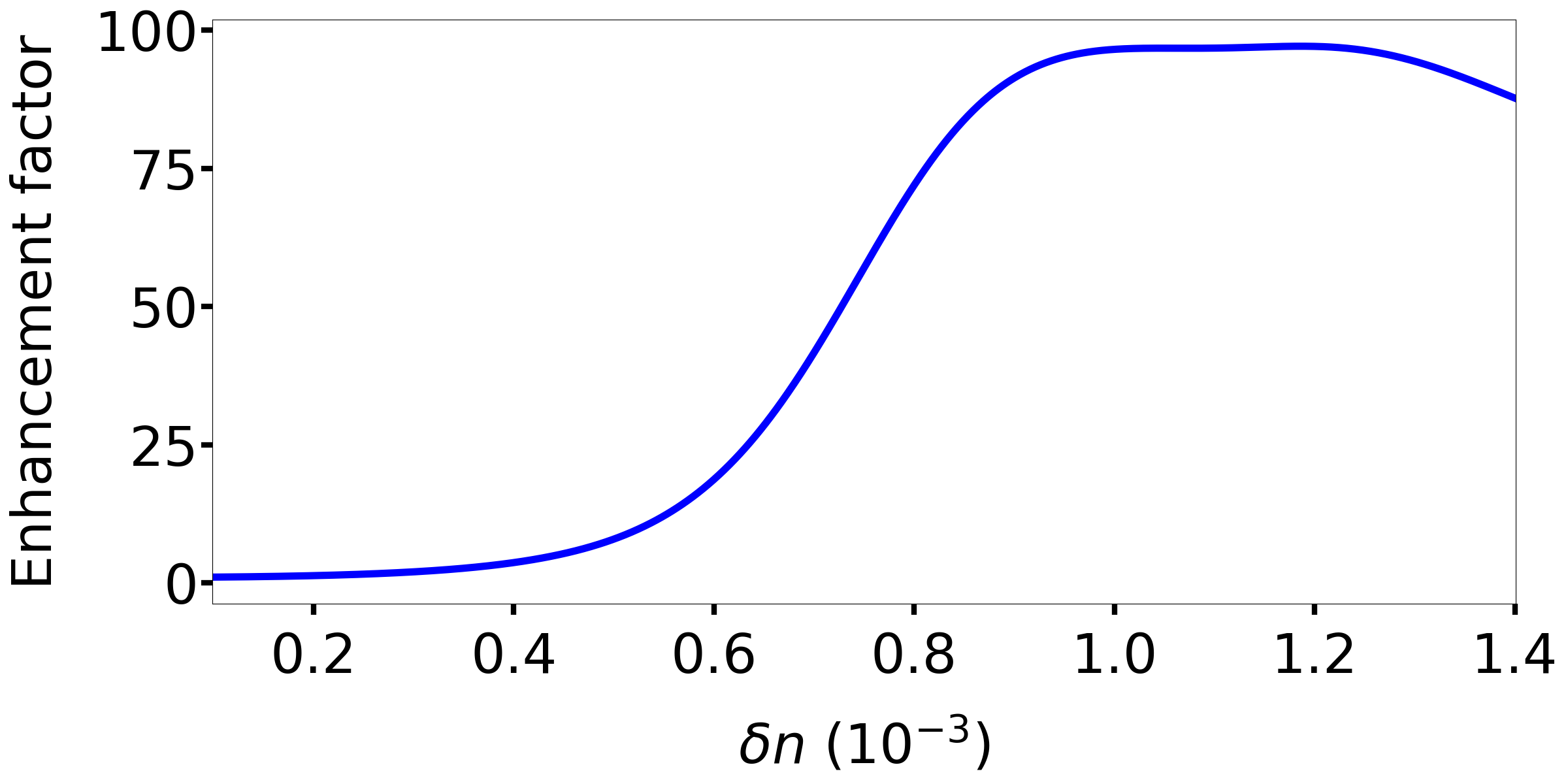}
		\caption{} 
        \label{fig:dn_enhancement}		
	\end{subfigure}
    \caption{(a) Output powers for fundamental and second harmonic modes versus grating strength $\delta n$. \\ (b)
    Corresponding SHG enhancement over a standard QPM device. $\Lambda_{s} = 5.78$\,mm, other parameters as in Fig.~\ref{fig:dsv}(b).}	
\end{figure}

The grating refractive index modulation amplitude $\delta n$ is ultimately what determines the coupling strengths of the slow-light grating and generates the enhancement. If we set $\delta n = 0$ we recover a standard QPM device and see no enhancement. Figure \ref{fig:dnv_ratios} shows the effect of varying the grating strength on the fundamental and harmonic output powers for an input intensity of $10^{3}$\,W/c$\text{m}^{2}$ and Fig.~\ref{fig:dn_enhancement} shows the corresponding increase in SHG efficiency compared to a standard QPM device.
At $\delta n = 10^{-4}$ the slow down factor is reduced to 1.01 resulting in almost zero SHG and therefore the device is behaving as a standard QPM device. As the grating strength is increased we see a corresponding increase in SHG which reaches a maximum conversion efficiency at $\delta n = 10^{-3}$. As $\delta n$ reaches $1.3\times10^{-3}$ the SHG begins to decrease with a corresponding increase in the backward fundamental mode. This is the same behavior we saw when increasing $\Lambda_{s}$. In both cases as the parameter is increased the field enhancement increases but if the field enhancement is increased beyond an optimum for SHG light is coupled into the backward fundamental mode.  

\section{\label{sec:level7}Conclusion}

In conclusion, we investigated the use of slow-light gratings in a quasi phase matched device for enhancing second harmonic generation. A phase-shifted superstructure grating creates a slow-light effect and leads to corresponding field enhancement which in turn enhances SHG. Since the superstructure grating couples the forward and backward propagating waves of the fundamental pump field, SHG also leads to forward as well as backward propagating harmonic fields. We therefore added a second, short Bragg grating at the start of the device to act as a reflector for the harmonic field, thus ensuring that all harmonic output is in the forward direction.

We found that for a given wavelength and input intensity there is an optimum superstructure period and thus an optimum group velocity reduction to maximize the conversion efficiency. The system also benefits from a resonator effect formed between the slow-light and the reflection grating and is therefore sensitive to the relative phase of the two gratings.

If the slow-light effect is too strong, for example because of a long superstructure period, a large refractive index modulation, or too high an input pump intensity, pump light starts to be back-reflected by the system and exits through the input port, thereby reducing the maximum achievable SHG efficiency. However, we found that the device still exhibits near-unity conversion efficiency for intensities spanning three order of magnitude. 

Most importantly, for all the parameter regimes investigated the slow-light device enhances significantly the SHG conversion efficiency compared to a standard quasi-phase matched device without the slow-light grating. For the realistic parameters of magnesium oxide doped thin film lithium niobate, enhancements by factors of several hundreds are predicted. Slow-light enhancement therefore allows for SHG at much shorter device lengths or at much lower pump intensities, which could have significant impact in low-power applications such as in quantum technology.


\begin{acknowledgements}
The authors acknowledge funding through the UK National Quantum Technologies Programme (EPSRC grant numbers EP/T001062/1, EP/M024539/1) and an EPSRC DTP PhD studentship. We also acknowledge the use of the IRIDIS High Performance Computing Facility, and associated support services at the University of Southampton, in the completion of this work.
\end{acknowledgements}


\bibliography{chi2enh}

\end{document}